\documentclass[twocolumn,reprint,aps,superscriptaddress,prl]{revtex4-1}

\usepackage{graphicx}
\usepackage{dcolumn}
\usepackage{bm}
\usepackage[utf8]{inputenc}
\usepackage{color}
\usepackage{amsmath}
\usepackage{amssymb}
\usepackage{amsbsy}
\usepackage{bbold}
\usepackage{graphicx}
\usepackage{newunicodechar}
\newunicodechar{₁}{$_1$}

\usepackage{physics}
\usepackage{textcomp}
\usepackage[]{todonotes}
\usepackage[unicode=true,
 bookmarks=false,
 breaklinks=false,pdfborder={0 0 1},backref=false,colorlinks=true]
 {hyperref}
\hypersetup{
 linkcolor=blue,citecolor=blue,urlcolor=blue,linkcolor=blue,citecolor=blue,urlcolor=blue}

\begin{document}

\preprint{APS/123-QED}

\title{Topological Braiding and Dynamic Probing of Phase Transitions at Temporal Interfaces in Non-Hermitian Synthetic Dimensions }

\author{Yuanhang Jiang}
\altaffiliation{These authors contributed equally to this work.}
\affiliation{School of Physics, Harbin Institute of Technology, Harbin 150000, People’s Republic of China}

\author{Jianfei Li}
\email{Corresponding author:jianfei\_li@hit.edu.cn}
\altaffiliation{These authors contributed equally to this work.}

\affiliation{School of Physics, Harbin Institute of Technology, Harbin 150000, People’s Republic of China}
\affiliation{Heilongjiang Provincial Key Laboratory of Plasma Physics and Application Technology, Harbin 150000, People’s Republic of China}

\author{Chengxi Yang}
\affiliation{School of Physics, Harbin Institute of Technology, Harbin 150000, People’s Republic of China}

\author{Ziyi Liu}
\affiliation{School of Physics, Harbin Institute of Technology, Harbin 150000, People’s Republic of China}

\author{Chen Chen}
\affiliation{School of Physics, Harbin Institute of Technology, Harbin 150000, People’s Republic of China}

\author{Hongyu Liu}
\affiliation{Department of Mathematics, City University of Hong Kong, Kowloon, Hong Kong SAR, China}

\author{Zhongxiang Zhou}
\affiliation{School of Physics, Harbin Institute of Technology, Harbin 150000, People’s Republic of China}
\affiliation{Heilongjiang Provincial Key Laboratory of Plasma Physics and Application Technology, Harbin 150000, People’s Republic of China}
\affiliation{Heilongjiang Provincial Innovation Research Center for Plasma Physics and Application technology, Harbin 150001, People's Republic of China}

\author{Jingfeng Yao}
\email{Corresponding author:yaojf@hit.edu.cn}
\affiliation{School of Physics, Harbin Institute of Technology, Harbin 150000, People’s Republic of China}
\affiliation{Heilongjiang Provincial Key Laboratory of Plasma Physics and Application Technology, Harbin 150000, People’s Republic of China}
\affiliation{Heilongjiang Provincial Innovation Research Center for Plasma Physics and Application technology, Harbin 150001, People's Republic of China}

\author{Chengxun Yuan}
\email{Corresponding author:yuancx@hit.edu.cn}
\affiliation{School of Physics, Harbin Institute of Technology, Harbin 150000, People’s Republic of China}
\affiliation{Heilongjiang Provincial Key Laboratory of Plasma Physics and Application Technology, Harbin 150000, People’s Republic of China}
\affiliation{Heilongjiang Provincial Innovation Research Center for Plasma Physics and Application technology, Harbin 150001, People's Republic of China}

\begin{abstract}

Non-Hermitian systems give rise to distinct topological phenomena, yet their manifestations at temporal interfaces characterized by abrupt changes in system parameters remain largely unexplored.
Upon an abrupt alteration of the Hamiltonian in a one-dimensional non-Hermitian system,the ensuring temporal interface excites both reflected and refracted wave modes.
By introducing a chiral-symmetric Hamiltonian, this study reveals the topological effects at such temporal interfaces.
We find that the reflection and refraction coefficients exhibit a topological braiding structure. This structure is directly determined by the difference in the topological invariants across the interface, establishing a bulk-boundary correspondence for temporal interfaces in non-Hermitian systems.
Furthermore, we propose a dynamical probe that leverages the geometric similarity of eigenstates at the temporal interface to detect topological phase transitions.
These findings establish a fundamental connection between topological braiding and nonreciprocal dynamics at temporal interfaces, providing a platform to explore phase transition detection and nonreciprocal phenomena in time-varying non-Hermitian systems.

\end{abstract}

\maketitle

\emph{Introduction---}Research in topological photonics has expanded from periodic arrangements of media in real space\cite{barsukova2024direct,benalcazar2017quantized} to synthetic dimensions\cite{lu2021topological,yang2022simulating,ehrhardt2023perspective,tang2022nonreciprocal,li2023direct}.
Synthetic dimensions represent a theoretical framework that utilizes internal degrees of freedom, such as frequency and angular momentum, to emulate additional spatial dimensions. 
Unlike lattices in real space, synthetic frequency lattices enable the direct construction of topological models incorporating next-nearest-neighbor or even long-range couplings, overcoming the limitation of nearest-neighbor interactions.
This provides a convenient approach for exploring higher-order topological phases and also makes it possible to study complex non-Hermitian systems featuring non-reciprocal couplings.
Recently, the topological braiding of energy bands has been experimentally demonstrated in non-Hermitian synthetic frequency lattices\cite{wang2021topological,zhang2023observation}.

The theory for manipulating topological phases in temporal photonic crystals via periodic modulation of the dielectric permittivity is well established\cite{yang2025topologically,lustig2018topological,zhou2020broadband},with associated topological boundary states emerging within momentum band gaps. Recently, exponential energy growth within momentum bandgaps has also been discovered\cite{lustig2023time}. The space-time duality of Maxwell's equations suggests that wave propagation at an abrupt temporal interface is analogous to reflection and refraction at a spatial interface between different optical media.
In contrast to the extensive research on temporal photonic crystals, the theory of temporal interface topology formed by different band topological structures is less developed.\cite{swanson2004transition,dong2024quantum,long2023time,li2025topological}. 
Extending the synthetic frequency lattice framework to temporal interfaces, recent work in Hermitian systems has demonstrated that the braiding topology of scattering coefficients at such interfaces can accurately map onto the system's global topological invariants\cite{xu2025probing}. 
However, when considering more complex non-Hermitian systems, a research gap remains in the theory of temporal interface topology. Since the eigenvectors of non-Hermitian systems are no longer orthogonal, the scattering coefficients at the temporal interface need to be redefined\cite{kawabata2019symmetry,zhang2025algebraic,zhang2022review,zhao2025observation}. 
This raises a new question: does the correspondence between the topological phases of the system across the temporal interface and the braiding of the scattering coefficients still hold? 

To address this challenge,  a theoretical framework is proposed based on a one-dimensional non-Hermitian Su-Schrieffer-Heeger (SSH) model\cite{geng2023separable,yao2018edge,zhang2022universal,li2020critical}. We introduce non-reciprocal intercell coupling and employ the biorthogonal basis to redefine the scattering coefficients, thus establishing a scattering theory for the non-Hermitian framework.
By modulating the coupling strengths in the synthetic frequency lattice, we configure non-Hermitian systems with distinct topological phases on either side of the temporal interface.  
This allows us to derive a temporal bulk-boundary correspondence for non-Hermitian systems.
We demonstrate that the braiding topology of the scattering coefficients remains a robust indicator of topological phases in non-Hermitian systems. This provides a method to detect topological phase transitions and determine the topological order, which we validate even in the presence of next-nearest-neighbor couplings\cite{geng2023separable}.
Moreover, we identify that the geometric similarity of eigenvectors across the temporal interface at specific wave vectors can serve as a direct indicator for topological phase transitions in non-Hermitian systems\cite{lieu2018topological,kleefeld2009construction,sheppard2025topological}. 
This finding stems from a synergistic effect between the intrinsic non-reciprocal gain mechanism of the non-Hermitian system and the modulation imposed by the temporal interface. 
We present the first theory of topological temporal interfaces in one-dimensional chiral-symmetric non-Hermitian systems, establishing a new perspective for wave manipulation in time-varying systems. \par
\emph{Topological effects at the temporal interface in non-Hermitian system ---} Considering two resonator rings with a coupling strength \( k \) and identical free spectral ranges \( \Omega \)[Fig.1(a)], their coupling gives rise to two new frequency modes,  \( l_1\Omega + k \) and \( l_2\Omega - k \), effectively constituting a one-dimensional synthetic frequency lattice\cite{li2023direct,zhang2023observation,xu2025probing}.
A phase modulator is embedded in one resonator ring to induce photon transitions, analogous to the one-dimensional Su-Schrieffer-Heeger (SSH) model. These transitions occur both within a unit cell and between adjacent unit cells.
By applying a modulation signal \( J(t) \) (Supplementary materials 3.1).
\begin{equation}
J(t)=4\sum_{l}\left[g_{l+1}\cos\left((l\Omega-2\kappa) t\right)+g_{l+1}^{\prime}\cos\left((l\Omega+2\kappa) t\right)\right]
\end{equation}
\begin{figure}[t]
\includegraphics[width=0.49\textwidth]{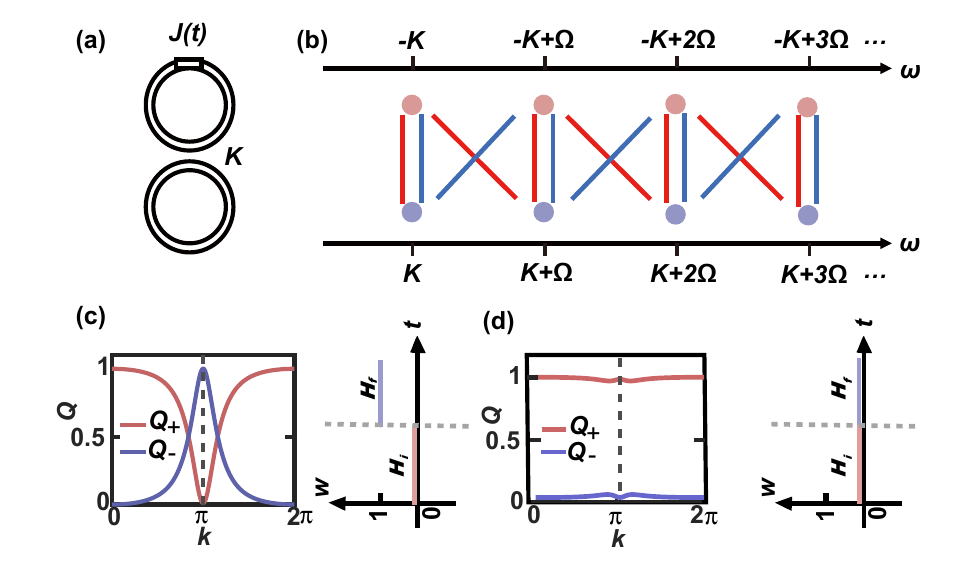}
\caption{\label{figs:fig1}  (a) Coupled resonator ring. (b) Synthetic frequency lattice formed by symmetric (pink dots) and antisymmetric (blue dots) modes. (c) Geometric similarity between the state vectors of the upper and lower bands after the temporal interface and the state vector of the upper band before the temporal interface, when the topological phases on the two sides differ. The reference coupling strength is denoted as\( g_0 = 1 \), the nonreciprocal coupling strength  \( \gamma = g_0/10 \).  Before the time interface, \( g_{1i} =g_0/2\), \( g_{2i} =g_0 \pm \gamma/2 \), After the time interface,  \( g_{1f} =2g_0\), \( g_{2f} =g_0 \pm \gamma/2 \) (d) Same as (c), but for the case where the topological phase remains unchanged across the temporal interface. Before the time interface, \( g_{1i} =3g_0/2\), \( g_{2i} =g_0 \pm \gamma/2 \), After the time interface,  \( g_{1f} =2g_0\), \( g_{2f} =g_0 \pm \gamma/2 \)}
\end{figure}
Coupling between different frequency modes is induced [Fig.1(b)].
Within this synthetic frequency lattice, we introduce the temporal interface by abruptly changing the coupling strengths between lattice sites at time \(t_0\).
Specifically, for \(t < t_0\), the system is described by coupling strengths \(g_{n i}\) and Hamiltonian \(H_{i}\).
At \(t = t_0\), the modulation signal is instantaneously switched, causing the coupling strengths to jump to \(g_{n f}\) and the Hamiltonian to change to \(H_{f}\).
This abrupt change of the Hamiltonian at \(t_0\) defines the temporal interface under investigation. 
The intercell coupling \( g_2 \) is nonreciprocal, while the intracell coupling \( g_1 \) is reciprocal. 
To quantify how the topological difference across the temporal interface affects the state vector, we introduce a physical quantity based on the biorthogonal basis: the geometric similarity, denoted \(Q_{\pm}\).
\begin{equation}
Q_{\pm}=\frac{|\langle L_{f}^{\pm} |R_{i}^{+}\rangle|^{2}}{\langle L_{f}^{\pm}| L_{f}^{\pm}\rangle\langle R_{i}^{+}|R_{i}^{+}\rangle}
\end{equation}
It measures the normalized overlap between the left post-interface eigenvector \(\langle L_f^{\pm}|\) of the system and the right pre-interface eigenvector  \(|R_i^{+}\rangle \).
This quantity effectively captures the geometric similarity of the system's eigenstates across the temporal interface. 
A change in the topological phase across the interface alters the global topological structure of the eigenvectors, which is reflected in distinct signatures in \(Q_{\pm}\).
The geometric similarity curves of the state vectors before and after the temporal interface are plotted in Fig. 1(c,d).
When the topological phases across the temporal interface remain consistent, the geometric similarity curves for the upper and lower bands diverge.
For wave vectors \(k\) across the entire Brillouin zone, the energy is predominantly concentrated in the transmitted waves associated with the upper band.
The relative magnitude between \(Q_{+}\) and \(Q_{-}\) is determined by (Supplementary materials 2.1):  
\begin{equation}
    \mathrm{sgn}(\frac{\sqrt{[(g_{1f} - g_{2f} )^{2} - \gamma_{f}^{2}/4][(g_{1i} - g_{2i} )^{2} - \gamma_{i}^{2}/4]}}{(g_{1i} - g_{2i} -\gamma_{i}/2)(g_{1f} - g_{2f}+\gamma_{f}/2)})
\end{equation}
The relative magnitudes of \(Q_{+}\) and \(Q_{-}\) serve as a probe for topological phase transitions since such transitions involve a band gap closure(Supplemental Material), where eigenvectors become highly sensitive and change drastically.
This reorganization directly affects the relation between \(Q_+\) and \(Q_-\), manifesting as an exchange of their magnitudes.
When the topological phases differ across the interface, the \(Q_{+}\) and \(Q_{-}\) curves intersect at \(k=\pi\).
This difference directly determines the critical relation between the intra-cell and inter-cell coupling strengths at the transition point: \(g_{1} = g_{2}-\gamma/2\).

We employ the biorthogonal basis method\cite{brody2013biorthogonal,lieu2018topological}, where \(|R_{i(f)}^{\pm}\rangle\) and \(\langle L_{i(f)}^{\pm}|\) denote the right and left eigenvectors before (after) the temporal interface. Within this framework, the refraction coefficients are defined as \(r_{+}=\langle L_{f}^{+} | R_{i}^{+} \rangle\) and \(r_{-}=\langle L_{f}^{-} | R_{i}^{+} \rangle\).
  At the Brillouin zone center \(k=\pi\), the overlap between pre- and post-interface states reveals the topological phase transition. If a transition occurs, the largest overlap is between the post-interface lower-band state and the pre-interface upper-band state. Conversely, in the absence of a transition, the post-interface state has a larger overlap with the pre-interface state of the same band (upper-upper) [Figs. 1(c), 1(d)]. 
  The relative magnitude of the geometric similarity curves, \(Q_{+}\) and \(Q_{-}\), at \(k=\pi\) serves as a signature for topological phase transitions, enabling the dynamical detection of temporal interface topology.
  When one of the off-diagonal elements of the effective Hamiltonian vanishes  \(E_{\pm} = \pm \sqrt{f(k) \cdot g(k)}=0\).
  By employing a nonreciprocal coupling via the signal \(J(t)\) , the energy gap closes at that point, resulting in a vanishing energy spectrum.
  we derive the effective Hamiltonian for a frequency lattice in the synthetic dimension, Since the Hamiltonian is non-Hermitian, where \(f(k)^* \neq g(k)\) (Supplementary materials 1.1).

\begin{small}
\begin{equation}
H = \begin{pmatrix}
0 & f(k) \\
g(k) & 0
\end{pmatrix}
\end{equation}
\end{small}

The topological braiding diagram of the temporal interface [Fig. 2(a), 2(b)] reveals that the braiding number equals the difference in the system’s topological phases across the interface. Here, the non-Hermitian winding number is defined as:\cite{lieu2018topological,zhou2019periodic,borgnia2020non,gong2018topological,altland1997nonstandard}:
\begin{equation}
w=\frac{i}{\pi}\int_{B Z}\left\langle\ L_{\pm}\left|\frac{\partial}{\partial k}\right| \ R_{\pm}\right\rangle d k
\end{equation}
Correspondingly, on the Bloch sphere, the eigenvector reaches either the South or North Pole when one of the off-diagonal elements of the Hamiltonian \(H(k) \) becomes zero\cite{lieu2018topological}.
As illustrated in Fig. 2, the topological braiding of the reflection and refraction coefficients on a torus directly manifests the difference in the system's topological invariant across the temporal interface. Specifically, the braiding number is 1 in Fig. 2(a), corresponding to a topological phase transition, whereas it is 0 in Fig. 2(b), where the topological phase remains unchanged. The trajectories of the Bloch vector for the upper and lower bands on the Bloch sphere are also plotted, providing a geometric visualization of the band topology. . 
The initial state before the interface is prepared with only the upper energy level excited. When a topological phase transition does not occur, the \(\boldsymbol{b_k}\) distribution curve fails to enclose the origin upon continuous variation. Owing to the chiral symmetry of the Hamiltonian, the state vectors of the upper and lower bands are distributed symmetrically with respect to the \(z\)-axis on the Bloch sphere (Supplementary materials 2.2).

\begin{figure}[t]
\includegraphics[width=0.49\textwidth]{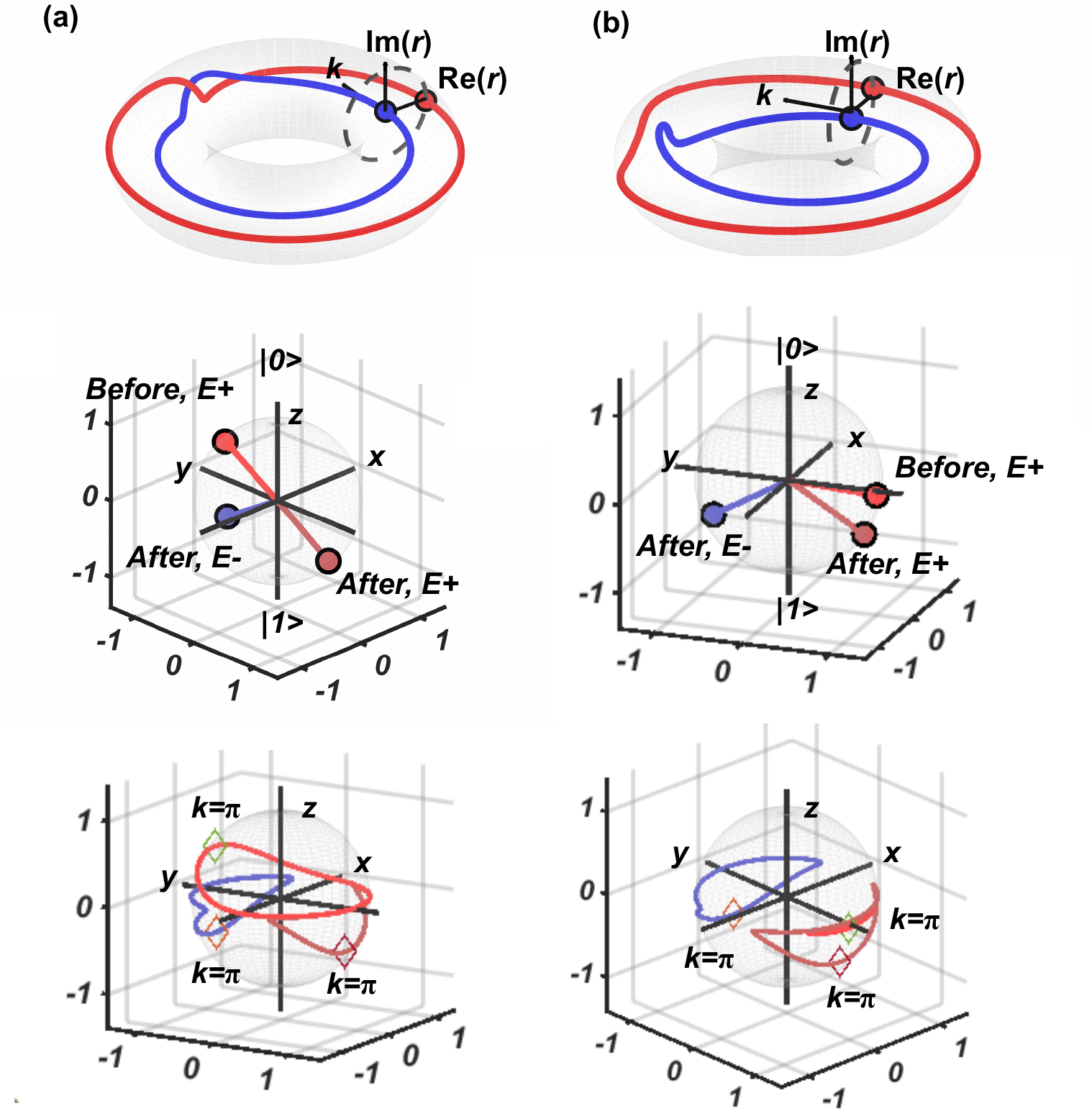}
\caption{\label{figs:fig2}  (a) Interwoven reflection (blue curve) and refraction (red curve) coefficients when topological phases differ across the temporal interface, plotted against the toroidal direction \(k\) and poloidal coordinate \(r\). On the Bloch sphere, red/blue shading indicates the distribution of the state vector and \(\boldsymbol{b_k}\) for the upper/lower band; poles are labeled with \(|0\rangle\) and \(|1\rangle\). The \(\boldsymbol{b_k}\) curve crosses a pole only at the topological phase transition.  Diamond symbols marking the transition points at the geometric similarity curves \(Q\)[see Fig. 1(c)] on the \(\boldsymbol{b_k}\) trajectory.  (b) Interwoven reflection (blue) and transmission (red) coefficients when topological phases are identical across the temporal interface.  }
\end{figure}

\begin{figure}[t]
\includegraphics[width=0.49\textwidth]{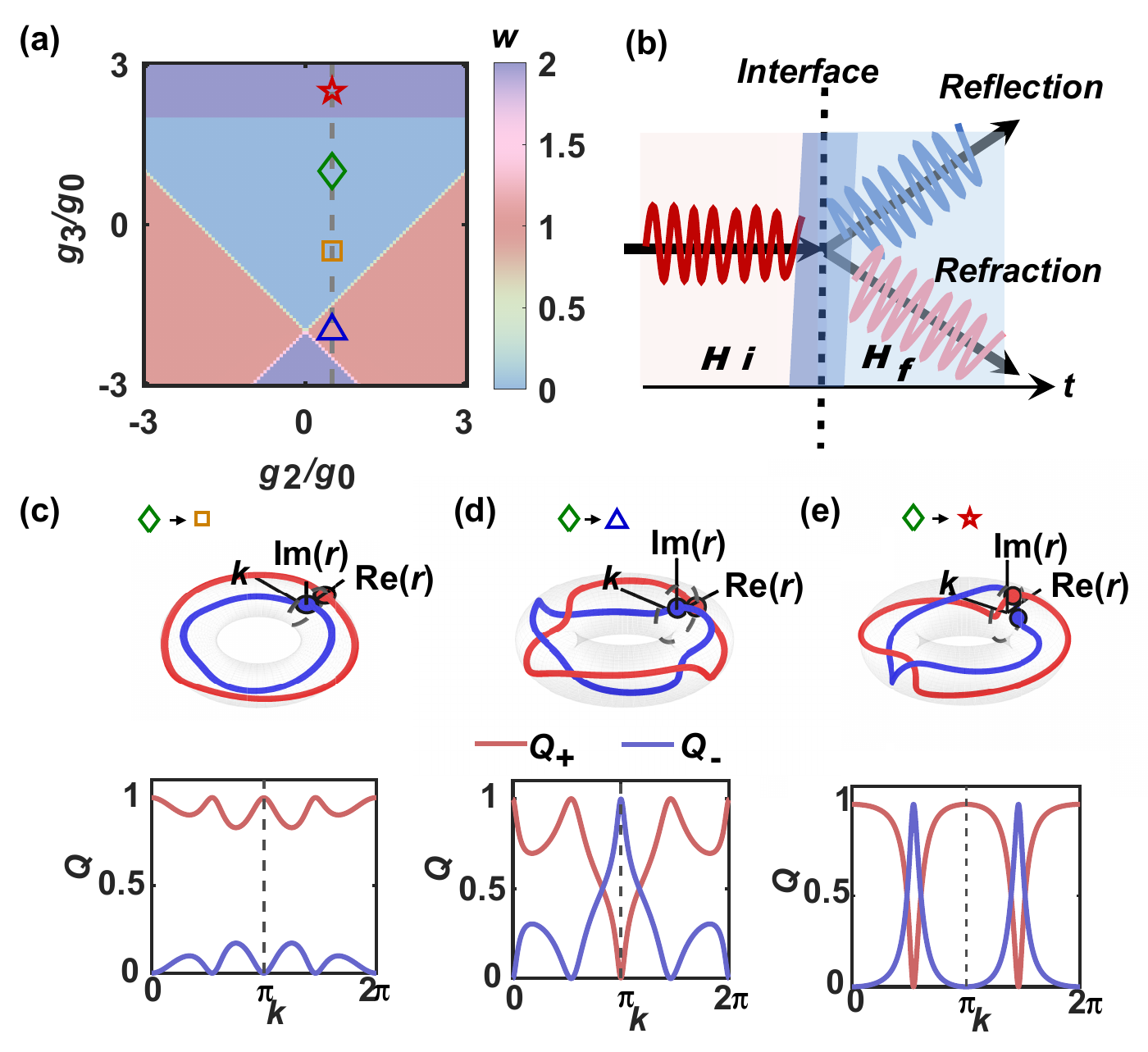}
\caption{\label{figs:fig3}Higher-order topological phase.  (a) Topological phase diagram with fixed intracell coupling strength \(g_1=2g_0\), intercell coupling strength \(g_2=g_0/2{\pm}\gamma/2\), and next-nearest-neighbor coupling strengths \(g_3=-2g_0\) (triangle), \(g_3=-g_0/2\) (square), \(g_3=g_0\) (diamond), and \(g_3=2.5g_0\) (star), corresponding to winding numbers of 1, 0, 0, and 2, respectively. (b) Schematic of the temporal interface effect. (c) The system undergoes no topological phase transition. (d) The winding number difference across the temporal interface is 1. (e) The winding number difference across the temporal interface is 2.    }
\end{figure}\par

To demonstrate the universality of topological braiding at temporal interfaces in non-Hermitian systems, we utilize a one-dimensional Su-Schrieffer-Heeger (SSH) model with nonreciprocal intercell coupling and next-nearest-neighbor interactions \cite{zhang2024topological,dias2022long,rafi2024exceptional}. Based on this model, we calculate the topological phase diagram[Fig. 3(a)].
The system remains topologically trivial before the temporal interface. By adjusting \(g_3\), the system enters distinct topological phases after the temporal interface. It is observed that the braiding number of the torus equals the topological difference across the temporal interface.
A comparison between the geometric similarity curve and the complex energy spectrum after the temporal interface reveals that the momentum points where the relative magnitudes of the geometric similarity switch correspond to those where the energy gap nearly closes. This correlation indicates that the evolution of the geometric similarity curve can signal imminent topological phase transitions.
Moreover, the number of points where the relative magnitude of geometric similarity changes matches the number of momentum points where the energy gap nearly closes.
After passing through the temporal interface, the system evolves from an initial Hamiltonian \(H_i\) to a final Hamiltonian \(H_f\), where the incident wave undergoes reflection and refraction[Fig. 3(b)].
At the temporal interface, the system's Hamiltonian changes from an initial \(H_i\) to a final \(H_f\). This temporal evolution scatters the incident wave into reflected and refracted components [Fig. 3(b)].
In the non-Hermitian system with nonreciprocal coupling, a comparison of Figs. 3(c)–3(e) confirms that the difference in the topological invariant across the temporal interface equals the winding number of the braiding pattern formed on a torus by the reflection and refraction coefficients.
In Fig. 3(c), where the system remains in a topologically nontrivial phase across the temporal interface, \(Q_{+}\) consistently exceeds \(Q_{-}\). As a result, after interacting with the temporal interface, the energy of the incident wave is predominantly concentrated in the refracted wave associated with the upper band.In Fig. 3(d), the geometric similarity curve shows that the relative magnitudes of \(Q_{+}\) and \(Q_{-}\) swap near \(k=\pi\).
In Fig. 3(e), \(Q_{+}\) and \(Q_{-}\) exchange their relative magnitudes at \(k\approx1.7\) and \(k\approx4.6\), which aligns with the bandgap closing points (Supplementary materials 2.1).

\emph{Propagation asymmetry across the temporal boundary---} In non-Hermitian systems, probability conservation is no longer universally upheld, due to the non-orthogonality of the Hamiltonian's eigenstates and the non-unitary nature of the time-evolution operator. This breakdown is particularly evident at a temporal interface, where the total probability before the interface generally differs from the sum of the probabilities of the transmitted and reflected waves after the interface. 
We define the projection coefficients as \(d_{+}= \langle L_{f}^{+} | \psi(0^{-}) \rangle\)and \(d_{-}= \langle L_{f}^{-} | \psi(0^{-}) \rangle\),The initial intensity \(I(0^{-})\)is then given by (Supplementary materials 3.2):
\begin{small}
\begin{equation}
\begin{split}
I(0^{-}) = & |d_{+}|^{2}\langle R_{f}^{+}|R_{f}^{+} \rangle + |d_{-}|^{2}\langle R_{f}^{-}|R_{f}^{-} \rangle \\
          & + d_{+}^{*}d_{-}\langle R_{f}^{+}|R_{f}^{-} \rangle + d_{-}^{*}d_{+}\langle R_{f}^{-}|R_{f}^{+} \rangle
\end{split}
\end{equation}
\end{small}

 This probability discontinuity stems from the distinct sets of eigenstates possessed by the Hamiltonians on either side of the interface. 
 When the initial state is projected onto the biorthogonal basis behind the interface, the non-Hermitian nature introduces non-orthogonalities between eigenstates of different energies, which act as sources for additional probability flux.
This temporal evolution is governed by the imaginary part of the energy spectrum \(\mathrm {Im}(E)\), which dictates exponential amplification or decay. In  contrast, at \(k=\pi\), the spectrum is purely real. Consequently, the probability exhibits an instantaneous discontinuity at the interface, which then remains time-invariant.
For other \(k\) points, however, the jump is followed by an exponential growth or decay dynamics. Within the biorthogonal framework, the total probability after the interface defined as \(\sum|\langle L_{f}^{\pm} | R_{i}^{+}(0^{+}) \rangle|^{2}\), is generally not conserved.
Crucially, because the system's time evolution is governed by a non-Hermitian Hamiltonian ,as seen in the equation \(I(t) = I(0) \exp( 2 \operatorname{Im}(E_n) t / \hbar)\), the resulting time-evolution operator becomes non-unitary\cite{xu2025probing}. This non-unitarity directly manifests as a non-conserved sum of probabilities for the refracted and reflected waves, which now exhibits a distinct dependence on the wave vector \(k\). 

\begin{figure}[t]
\includegraphics[width=0.49\textwidth]{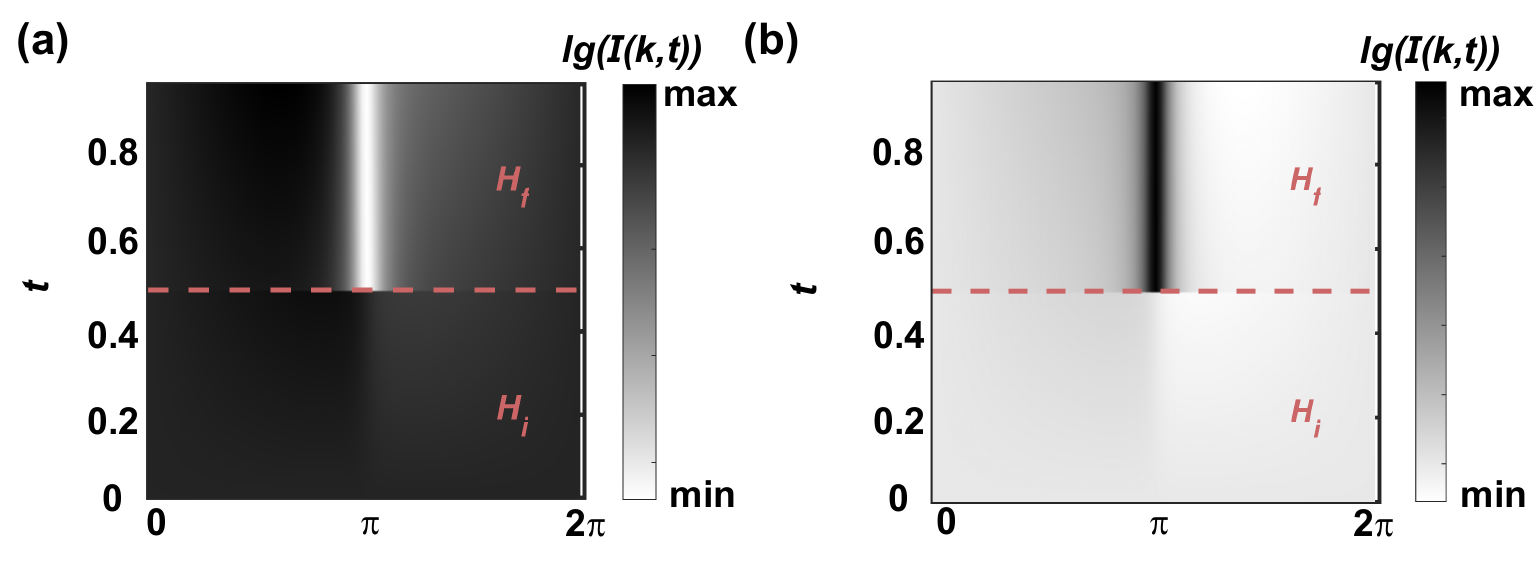}
\caption{\label{figs:fig4} Heatmaps of wavefunction intensity evolution across a temporal interface. Color bars represent the base-10 logarithm of the wavefunction intensity modulus squared. The horizontal and vertical axes correspond to the Brillouin zone wavevector and time, respectively. \(g_2\) remains constant across the temporal interface, where \(g_2=g_0\pm0.05g_0\), with next-nearest-neighbor and longer-range couplings neglected. (a) Forward transmission with \(g_1\) switching from 1.2\(g_0\) to 0.8\(g_0\) at the temporal interface. (b) Reverse transmission, complementary to (a), with \(g_1\) switching from 0.8\(g_0\) to 1.2\(g_0\). }
\end{figure}

The heatmaps [Fig. 4] validate the above conclusions by illustrating the wavefunction dynamics across the temporal interface.
For the forward transmission case[Fig. 4(a)], where \(g_1\) switches from 1.2 to 0.8, the wavefunction intensity shows a distinct attenuation near \(k=\pi\) at the temporal interface.
The intensity after this jump remains constant in time, consistent with the theory that the imaginary part of the energy spectrum \(\text{Im}(E)\) is zero at this point.
In regions where \(k\) deviates from \(\pi\), the jump is followed by either exponential growth (bright regions) or decay (dark regions) of the intensity, reflecting the non-unitary time evolution caused by a non-zero \(\text{Im}(E)\).
Conversely, the reverse transmission process in Fig. 4(b) exhibits a pronounced enhancement near \(k=\pi\).
These results demonstrate the non-reciprocal nature of the temporal interface effect: the transmission characteristics depend strongly on the direction of the parameter change.

\emph{Conclusion---} This work investigates topological effects at temporal interfaces within a synthetic frequency dimension, under the framework of non-Hermitian topological photonics.
We redefine the scattering coefficients at the temporal interface based on the biorthogonal basis method to describe the interface process when the Hamiltonian undergoes an abrupt change.
A fundamental mechanism in non-Hermitian topological photonics is unveiled in this work, through the establishment that the braiding topology of scattering coefficients at a temporal interface provides a direct and robust measure of the difference in topological invariants across the interface, even in the presence of nonreciprocal couplings. The geometric similarity of eigenstates at the temporal interface can also predict topological phase transitions and the number of transition points, which provides a new perspective for dynamically detecting topological phase transitions in systems. Moreover, temporal refraction and reflection waves exhibit novel controllable phenomena. For instance, the topological phases on both sides of the temporal interface can be leveraged to redetermine the energy distribution between the temporal refraction and reflection waves. Non-reciprocal transmission can also be achieved by controlling the direction of parameter changes.
The key advantage of this scheme is its superior experimental feasibility. Implementing a temporal interface in synthetic dimensions solely through coupling-strength modulation is far simpler than fabricating temporal photonic crystals that require periodic dielectric constant modulation. This advantage establishes a solid foundation for advanced studies into multi-layer temporal interface modulation,nonlinearity\cite{cai2025versatile,Yoshida2024ExceptionalPO,xia2020nontrivial}, and non-Hermitian topology\cite{Galiffi2021PhotonicsTI,wen2022unidirectional,Yan2023AdvancesAP}.

\emph{Acknowledgments}. This work is supported by the National Natural Science Foundation of China (NSFC, Contracts No. 12575213, 12505229,12505230 and U2541210), Heilongjiang Province Young Science and Technology Talent Support Program (No.2023QNTJ016), Natural Science Foundation of Heilongjiang Province of China (No.YQ2024A008), the China Radio Wave Propagation Research Institute Stable Support for Scientific Research Funding under Project (A251200020) and National Key R\&D Program of China (No.2025YFF0512000).

\bibliographystyle{unsrt}
\bibliography{sample}

\end{document}